\documentclass{article}[12pt]
\usepackage[capposition=top]{floatrow}
\usepackage[utf8]{inputenc}
\usepackage[margin=1.35in]{geometry}
\usepackage{authblk}
\usepackage{hyperref}
\usepackage{graphicx}
\usepackage{float}
\usepackage{subcaption}
\usepackage{amsmath}
\usepackage{graphics}
\usepackage{amsfonts}
\usepackage{comment}
\usepackage{apacite}
\usepackage{ragged2e}
\usepackage{textcomp}
\usepackage[labelfont=bf]{caption}
\usepackage[font={small,it}]{caption}
\usepackage{floatrow}
\usepackage[table]{xcolor}
\usepackage{hhline}
\usepackage[export]{adjustbox}
\RequirePackage{filecontents}
\usepackage{tabularx}
\usepackage{lscape}
\usepackage{enumerate}
\usepackage{rotating} 
\usepackage{xcolor}
\usepackage{stackengine}
\newlength\cellv
\newlength\cellh
\newlength\vertoffset
\newlength\horzoffset
\usepackage{tikz}
\usetikzlibrary{matrix}
\usetikzlibrary{arrows}
\usetikzlibrary{decorations.markings}

\usepackage{dsfont}

\title{DebtStreamness: An Ecological Approach to Credit Flows in Inter-Firm Networks}
\author{Anah\'i Rodr\'iguez-Mart\'inez$^{1}$, Silvia Bartolucci$^1$, Francesco Caravelli$^2$,\\ Victoria Landaberry$^3$, Pierpaolo Vivo$^4$, Fabio Caccioli$^{1,5,\star}$}

\date{ \small $^1$ Dept. of Computer Science, University College London,  WC1E 6EA London (UK) \\  
$^2$Theoretical Division,
Los Alamos National Laboratory, Los Alamos, New Mexico 87545 (USA)\\$^3$ Banco Central del Uruguay, 777 Diagonal J.P. Fabini 11100 Montevideo (Uruguay)\\$^4$ Dept. of Mathematics, King's College London, Strand WC2R 2LS London (UK)\\$^5$  Systemic Risk Centre, London School of Economics and Political Sciences, WC2A 2AE, London (UK)\\ $^\star$ Corresponding author: \url{f.caccioli@ucl.ac.uk}
}

\begin{document}

\maketitle

\begin{abstract}
Understanding how credit flows through inter-firm networks is critical for assessing financial stability and systemic risk. In this study, we introduce \textit{DebtStreamness}, a novel metric inspired by trophic levels in ecological food webs, to quantify the position of firms within credit chains. By viewing credit as the ``primary energy source" of the economy, we measure how far credit travels through inter-firm relationships before reaching its final borrowers. Applying this framework to Uruguay’s inter-firm credit network --- using survey data from the Central Bank --- we find that credit chains are generally short, with a tiered structure in which some firms act as intermediaries, lending to others further along the chain. We also find that local network motifs such as loops can substantially increase a firm’s DebtStreamness, even when its direct borrowing from banks remains the same. Comparing our results with standard economic classifications based on input-output linkages, we find that DebtStreamness captures distinct financial structures not visible through production data. We further validate our approach using two maximum-entropy network reconstruction methods, demonstrating the robustness of DebtStreamness in capturing systemic credit structures. These results suggest that DebtStreamness offers a complementary ecological perspective on systemic credit risk and highlights the role of hidden financial intermediation in firm networks.
\end{abstract}

\section{Introduction}

In any economy, the production dynamics of goods and services depend on the complex relationships between firms. Often, the output of one firm is linked to the products or services of another. Every sector requires specific inputs to manufacture its goods, and then supplies those goods to other sectors to fulfill their specific needs \cite{Carter1952}. Input-output networks \cite{Leontief1951} illustrate how raw materials flow from primary sectors --- like agriculture and mining --- to manufacturing sectors for processing, and finally to service sectors.

This is akin to how nutrients flow through a food web, from primary producers (such as plants and phytoplankton) to various levels of apex consumers, such as herbivores and carnivores, sustaining the complex ecosystems observed in nature \cite{ulanowicz2012growth}.

Both ecology and economics use network representations to capture the flow of resources, goods, or influence within such complex systems \cite{montoya2006ecological, demirel2022network, orenstein2021changing}. To better understand the role of individual agents --- whether species in an ecosystem or sectors in an economy --- similar metrics have been developed to quantify their position within these networks.

In ecological food webs, species are connected by predator-prey relationships. A species’ position within this web is captured by its \textit{trophic level}, indicating the relative
position that it occupies in the ecosystem and their distance from primary energy sources,  i.e.,
apex predators have a larger trophic level than phytoplankton \cite{hannon1973structure, levine1980several, scotti2009using}.

In economics, DownStreamness and UpStreamness measure a sector's position with respect to primary factors of production and final consumers, respectively \cite{bartolucci2025upstreamness}. Specifically, UpStreamness measures the distance of a firm or sector from final demand: an upstream sector sells a small share of its output to final consumption and a large share to other sectors for producing intermediate output \cite{Antras2012}. Similarly, DownStreamness represents the distance of a given sector from the economy’s primary factors of production \cite{Miller2015}. These measures provide a clearer picture of how many intermediate stages a product or service passes through before reaching the final consumer, although the two measures have a complex and far from intuitive interplay \cite{Silvia2023}.

In both ecological and economic networks, the structure of interactions and the position of nodes within the network are closely linked to the system’s stability and its response to exogenous shocks \cite{mcnerney2022production,haldane2011systemic, Battiston2012}. May’s pioneering work on random ecosystems \cite{may1972will} highlighted how connectivity influences stability, providing a framework to discuss the observed robustness of real systems. For instance, \shortciteA{dunne2002network} demonstrated that food web stability is influenced by the non-random distribution of interaction strengths, showing resilience when strong and weak links coexist. Similarly, \shortciteA{neutel2002stability} found that food webs are organized such that long loops contain many weak links, which critically contributes to their stability. \shortciteA{allesina2009googling} further emphasized that a species' position within the network, rather than its number of connections, is key to understanding its impact on co-extinctions, using an algorithm to identify species critical to food web stability.

In economics, Leontief’s foundational analysis showed that the response of an economy to a demand shock is determined by the structure of the input-output network through the so-called Leontief inverse, which accounts for the paths through which the shock propagates between sectors \cite{Leontief1951}. More recently, \shortciteA{acemoglu2012network} showed how aggregate macroeconomic fluctuations can emerge from such production networks, while \shortciteA{Elliott2022} argued that supply networks with intermediate productivity levels are particularly fragile in equilibrium. The complexity of supply chains became especially evident during the COVID-19 pandemic, which exposed the vulnerability of many firms. For instance, according to the US Census Small Business Pulse Survey (2020), $36\%$ of small businesses reported delays with domestic suppliers—particularly in the manufacturing, construction, and trade sectors—causing cascading disruptions in industrial supply chains \shortcite{Helper2021}.

The ecological perspective has recently gained traction in the study of complex economic and financial systems, where the concept of trophic levels is being adapted and enriched to understand structural stability and risk propagation. For example, \shortciteA{sansom2021trophic} introduced the notion of trophic coherence --- a measure of how neatly hierarchical a network is --- as a key determinant of stability in economic and financial networks. Their work highlights that systems with higher trophic coherence tend to be more resilient to shocks, echoing patterns observed in ecological food webs. Beyond this, analyses of firm-level hierarchies and network topology have revealed that structural complexity itself can be a source of systemic risk, with certain configurations amplifying vulnerabilities within financial institutions and markets \cite{LUMSDAINE2021100804}. These insights underscore the relevance of ecological frameworks in uncovering hidden fragilities in economic and financial networks.

However, existing economic analyses largely focus on material flows --- neglecting the equally critical pathways of financial flows. Moreover, while trophic concepts have begun to inform studies of supply chains and financial systems, a systematic framework to quantify positions within credit networks remains absent.

In this study, we bridge this gap by introducing \textit{DebtStreamness} --- a novel metric inspired by trophic levels --- to quantify a firm's position within credit chains. Firms extend credit to each other based on various operational needs, and these credit relationships form a directed network of inter-firm obligations. The DebtStreamness metric measures the distance separating a firm from the banking sector, which we consider the originator of credit in the system.  When efficiently allocated, credit---acting as the primary ``energy'' source of the economy---flows through the network, fueling productivity and enabling economic activity.  
By analyzing DebtStreamness, we gain a structured understanding of how credit flows from its primary source (the banking sector) to the various firms that make up the economy. This approach not only captures the length of credit chains, but also offers insights into potential bottlenecks and hidden financial intermediaries that could propagate or absorb exogenous and endogenous shocks.

We apply this framework to the inter-firm credit network of Uruguay, leveraging a unique dataset collected by the Central Bank of Uruguay during a 2018 survey of the country's 240 largest firms \cite{LANDABERRY2021100032}. While the dataset is partial --- each firm reports only its top three creditors ---it nonetheless provides rare, granular insight into the structure of inter-firm credit relationships. To assess the reliability of our results given this limitation, we complement our analysis with two maximum-entropy network reconstruction techniques, incorporating additional survey data on total inter-firm credit exposure. This approach allows us to evaluate whether information limited to a firm's top creditors is sufficient to accurately capture its position within the credit network as measured by our DebtStreamness metric, thereby testing the robustness and practical applicability of our methodology.

Our analysis reveals a tiered credit structure in Uruguay’s inter-firm credit network: a group of firms borrows primarily from the banking sector and, in turn, provides credit to others. We also find a strong negative correlation between the share of a firm's credit sourced directly from banks and its DebtStreamness. This finding is intuitive --- greater reliance on bank credit leads to shorter credit chains. However, we also show that network-specific features can significantly affect a firm's DebtStreamness, highlighting the complex interplay between local structure and systemic credit flows. We also show that sector-level aggregation can obscure these patterns, emphasizing the importance of granular data. 

By comparing our results with a standard upstream/downstream classification of firm types based on production network analysis, we demonstrate that inter-firm credit analysis provides complementary insights beyond what production-based approaches capture.

By offering a fresh lens on credit flows, our work provides complementary insights to traditional production-based analyses and highlights firms whose network positions may render them critical to financial stability, regardless of their size. This interdisciplinary framework opens new avenues for policymakers, regulators, and researchers aiming to monitor systemic risk in increasingly interconnected economies.

The paper is organized as follows: Section \ref{sec:method} introduces the definition of DebtStreamness and describes the dataset used in our analysis. Section \ref{sec:results} presents our results at the level of the granular inter-firm network, the sectoral level, and includes robustness checks based on partially reconstructed networks. Finally, Section \ref{sec:conclusions} concludes with a discussion of our main findings and their implications.

\section{Methods} \label{sec:method}
In this section, we present the DebtStreamness metric for measuring firms’ positions within inter-firm credit networks, describe the dataset used in our analysis, and outline the network reconstruction methods employed to address data incompleteness and validate the robustness of the methodology.

\subsection{DebtStreamness: Measuring Positioning in Credit Networks}
To capture the flow of credit within an economy, we model the inter-firm financial system as a \textit{directed, weighted network} of $N$ nodes, where each node represents a firm and each link denotes a credit relationship. Specifically, the network is encoded in a matrix $L$, where each element $L_{ij}$ reflects the amount borrowed by firm $i$ from firm $j$. We further denote by $D_i$ the total amount of borrowing of firm $i$, and by $B_i$ the amount borrowed by firm $i$ from the banking sector.

We focus exclusively on credit originating from \textit{banks} or \textit{other firms} within the inter-firm network, omitting alternative sources of credit such as bonds, private equity, or government loans. This assumption leads to the following accounting identity for each firm:

\begin{equation}\label{eq:identityTotalDebt}
D_i = B_i + \sum_{j=1}^N L_{ij} \ .
\end{equation}

Our objective is to quantify a firm's position within the credit network, interpreted as its average distance from the banking sector --- the primary originator of credit, analogous to primary producers in ecological food webs. This ``distance" reflects how many layers of inter-firm credit a unit of currency traverses before reaching a given firm. The banking sector serves as a natural reference point for measuring distance within the network.

To formalize this, we adapt the concept of DownStreamness  --- originally developed for input-output production networks \cite{Antras2012, Miller2015} --- to the context of inter-firm credit flows. 

Following \shortciteA{Antras2012}, we therefore define the DebtStreamness $DS_i$ of firm $i$ as a weighted sum over all possible paths from the banking sector to firm $i$. Each path is weighted by its length (number of intermediaries) and by the fraction of credit transmitted along that path from the banking sector to firm $i$.

Let $\ell_{ij} = L_{ij} / D_j$ represent the fraction of firm $j$'s debt that flows to firm $i$. Then, the DebtStreamness of firm $i$ is given by:

\begin{equation}\label{eq:DebtStreamnessDistanceSum}
 D S_i={\frac{B_i}{D_i}} + 2 {\sum_j \frac{\ell_{i j} B_j}{D_i}}+ 3 {\sum_{j k} \frac{\ell_{i k} \ell_{k j} B_j}{D_i}}+\ldots \ .
 \end{equation}

\begin{figure}[H]
    \centering
    \caption{\textbf{DebtStreamness (DS) of firm $i$.} This diagram shows the gross debt value of firm $i$, positioned downstream in the final stage of the debt chain within the production process, relative to banks, which are upstream and act as the credit originators. We show three different toy scenarios from the firm $i$'s perspective. Scenario I refers to the situation when firm $i$ borrows from the bank (initial node). Scenario II reflects the scenario where $i$ borrows from $j$, which in turn borrows from the bank. Finally, scenario III refers to the situation where $i$ borrows from $k$, which borrows from $j$, which borrows from the bank (initial node).}
    \label{fig:DSscheme}
    \includegraphics[width=\textwidth]{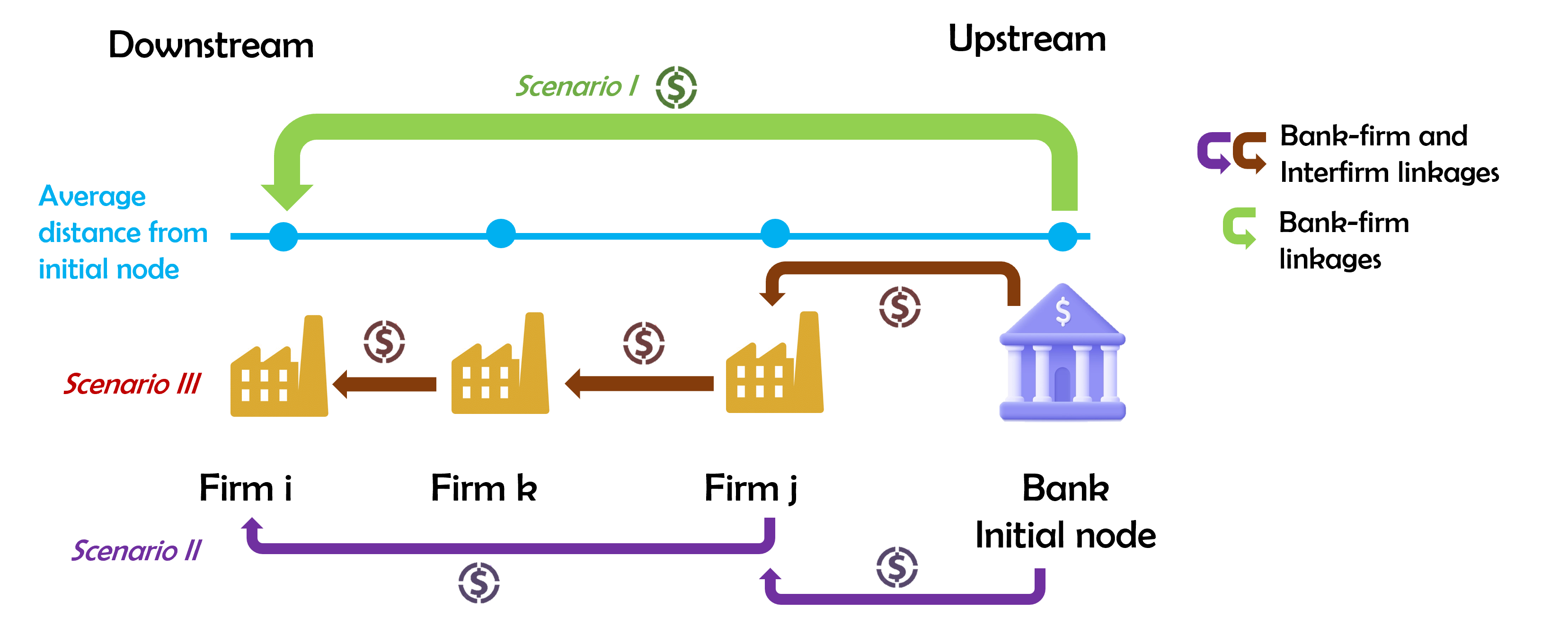}
\end{figure}

On the right-hand side of the above equation, the first term accounts for the direct borrowing of firm $i$ from the banking sector, the second term for the borrowing that is associated with credit chains or paths of length 2 (i.e., borrowing that is intermediated by one firm), the third term for paths of length $3$ (intermediated by two firms), and so on.
In Figure \ref{fig:DSscheme} we provide a graphical illustration of the first three terms of the equation above, corresponding to different borrowing scenarios and length of credit chains.

Following \cite{Antras2012}, it can be shown (see Appendix \ref{app:DebstDerivation}) that the above definition is equivalent to the following expression:

\begin{equation}\label{eq:DebtStreamnessDistance}
 D S_i=1 + \sum_j A_{ij} DS_j \ ,
 \end{equation}
 where $A_{ij}=L_{ij}/D_i$ denotes the share of firm $i$'s debt that belongs to firm $j$. 

This mirrors the definition of \textit{trophic levels} in ecology \cite{Silvia2023}: just as a species' trophic level equals one plus the average trophic level of the species it feeds on, the DebtStreamness of a firm equals one plus the weighted DebtStreamness of its creditors.

If we consider all nodes in the network $i=1,
\ldots, N$ and collect the DebtStreamness of all nodes in a vector $\overrightarrow{DS}$, \eqref{eq:DebtStreamnessDistance} can be written as
\begin{equation}
 {\overrightarrow{DS}}=1 + A~\overrightarrow{DS}\ ,
\end{equation}
with solution
\begin{equation}
 \overrightarrow{DS}=\left(\mathds{1} -  A\right)^{-1}\overrightarrow{1}\ ,
\end{equation}
where $\mathds{1}$ is the identity matrix and $\vec{1}$ a vector with all entries equal to one.

\subsection{Data}
We apply this framework using a unique dataset from the \textit{Economic Expectations Survey} conducted by the Central Bank of Uruguay in 2018~\cite{LANDABERRY2021100032}. The survey includes detailed credit relationship data for $240$ large firms, each with over $50$ employees.  The survey excludes sectors such as primary industries, financial intermediaries, the public sector, and real estate.

For each firm, the survey provides monthly information about:
\begin{itemize}
    \item Total commercial debts and sales credit,
    \item Its \textit{top three creditors and debtors}, along with corresponding amounts borrowed from and lent to them.
\end{itemize}

By incorporating both surveyed firms and their reported counterparties (i.e., top creditors and borrowers, many of which were not directly surveyed), we obtain an inter-firm credit network of $1,072$ firms. Although partial --- as surveyed firms disclose only their top creditors and borrowers, with even less data for non-surveyed firms ---  this dataset provides rare, granular insight into credit structures.  We have also conducted analyses using reconstructed networks, as discussed in Sec. \ref{sec:reconstructionmeth}, \ref{sec:reconstruction}, to confirm the robustness of our findings.

The data obtained through the survey is, then, complemented by the Central Bank Credit Registry database, which contains precise information about the borrowing of firms from banks, and which we use to compute $B_i$ for each firm $i$. Some firms in the dataset lack incoming paths from the banking sector. Consequently, their DebtStreamness is undefined, and they do not contribute to the DebtStreamness of other nodes in the system. These nodes have, therefore, been excluded from the analysis. 

\begin{figure}[H]
     \centering
\includegraphics[width=0.8\textwidth]{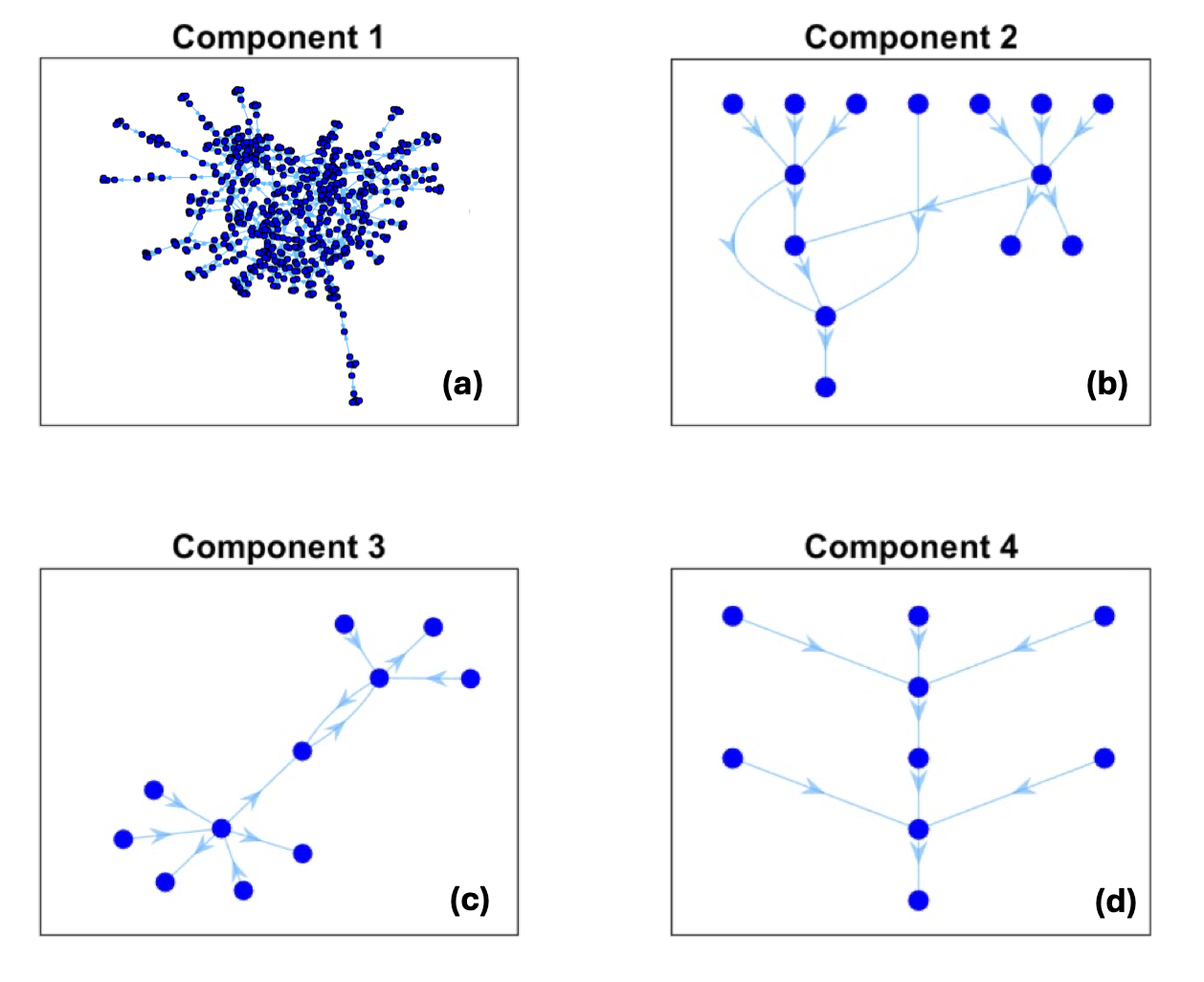}
      \caption{Main connected components of the inter-firm credit network.}   \label{fig:networkRepresentation}
\end{figure}

The resulting network comprises $843$ nodes, and it is made of several components. In Figure \ref{fig:networkRepresentation}, we show four of these components, including the largest one (Fig. \ref{fig:networkRepresentation}, panel $(a)$) with $675$ nodes. 
 
\begin{figure}[H]
    \centering
    \caption{Distribution of inter-firm credits. The data are well-fitted by a log-normal distributions of parameters $\mu\approx 11$ and $\sigma\approx 2$.}
    \label{fig:distributionEntries}
    \includegraphics[width=10cm]{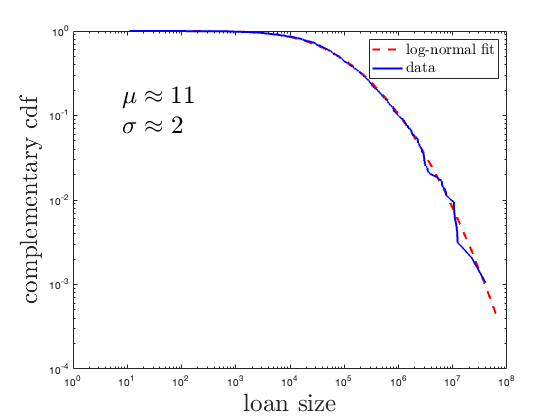}
\end{figure}

The distribution of inter-firm credits is well-fitted by a log-normal distributions of parameters $\mu\approx 11$ and $\sigma\approx 2$ as shown in Fig. \ref{fig:distributionEntries}. The properties of our networks are compatible with typical observed behaviour in inter-firm networks \cite{bacilieri2023firm}.

\begin{figure}[H]
    \centering
    \caption{Fraction of inter-firm credit associated with the top three creditors of surveyed firms. The distribution is quite heterogeneous, but on average the top three creditors of each firm account for about $50\%$ of their inter-firm credit. }  
    \includegraphics[width=0.7\textwidth]{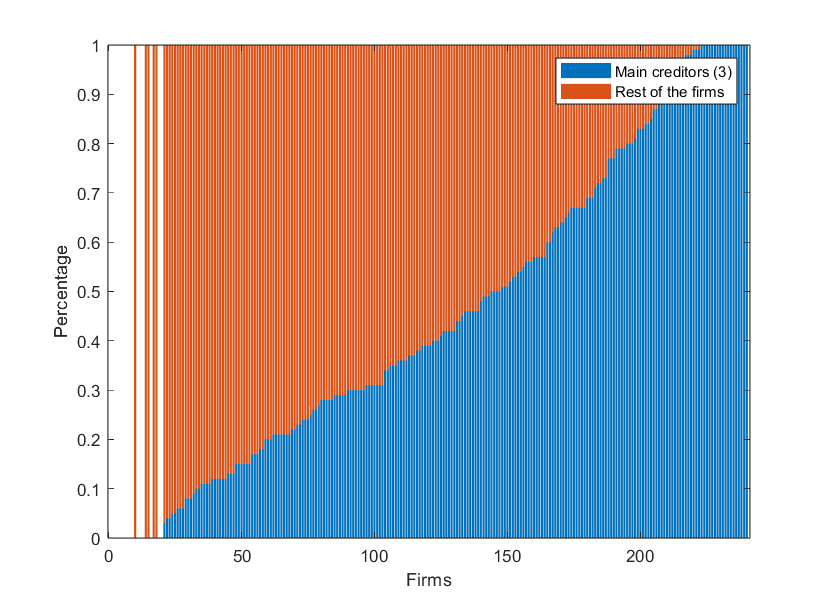}
    \label{fig:3firms}
\end{figure}

In Figure \ref{fig:3firms}, we report the share of total inter-firm credit accounted for by the top three creditors of each firm. While, on average, the top three creditors represent about $50\%$ of a firm's inter-firm credit, the distribution is highly heterogeneous. This raises the question of how the residual credit---i.e., credit relationships beyond the top three---might affect our results. To address this, we conduct robustness checks on partially reconstructed networks that incorporate this additional information as discussed in Sec. \ref{sec:reconstructionmeth}.

\subsection{Network reconstruction and robustness analysis} \label{sec:reconstructionmeth}

In financial network analysis, it is common to work with incomplete data, as full bilateral exposures between institutions are often unavailable due to privacy, reporting thresholds, or survey limitations \cite{Bacilieri2023,ialongo2022reconstructing}. To overcome this, network reconstruction methods—particularly those based on maximum entropy or sparsity assumptions—are widely employed to estimate the missing entries in the exposure matrix while preserving key known constraints \cite{cimini2021reconstructing}.

In our case, the empirical dataset provides partial network information: for each surveyed firm, the top three creditors and debtors are disclosed. However, the total volume of inter-firm credit for each firm is also reported. This allows us to test the robustness of our DebtStreamness measure under different plausible reconstruction methods of the entire original credit network.

In particular, we denote by $L^{\text{top}}$ the observed inter-firm credit matrix, containing only the top three credit relationships for each surveyed firm. Let $F_i$ denote the total volume of inter-firm credit borrowed by firm $i$, as reported in the survey. The residual, or unobserved, portion of credit for firm $i$ is then given by:
\begin{equation}
    R_i = F_i - \sum_j L^{\text{top}}_{ij} \ .
\end{equation}

To explore how this residual credit may influence DebtStreamness values, we construct two alternative partially reconstructed networks that represent two extreme scenarios of possible credit allocation.

\paragraph{Fully Connected Network Reconstruction.} In this case, we assume that the residual credit of each firm is uniformly distributed among all other firms not already listed as one of its top three creditors, hence filling uniformly  all zero entries of the $i$-th row of $L^{\text{top}}$. Formally:

\begin{equation}
    L^{\text{full}}_{ij} = 
    \begin{cases}
    L^{\text{top}}_{ij}, & \text{if } L^{\text{top}}_{ij} > 0, \\
    \dfrac{F_i - \sum_k L^{\text{top}}_{ik}}{N - 3} = \dfrac{R_i}{N - 3}, & \text{if } L^{\text{top}}_{ij} = 0,
    \end{cases}
\end{equation}
where $N$ is the total number of firms in the network.

\paragraph{Sparse Network Reconstruction.} Here, we aim to preserve the sparsity of the observed network. For each firm $i$, we determine the minimum number $N_i$ of additional creditors such that if $R_i$ is evenly distributed among them, each entry remains smaller than the smallest value in $i$'s top three observed credit links. This prevents the reconstructed links from dominating the observed structure. We then randomly assign the residual credit evenly across $N_i$ randomly selected zero entries in row $i$ of $L^{\text{top}}$.

These reconstruction methods allow us to test whether DebtStreamness is sensitive to the unknown distribution of the remaining credit. In Section~\ref{sec:reconstruction}, we compare the DebtStreamness values computed from $L^{\text{top}}$ with those obtained using the two reconstructed networks. The resulting high correlations confirm that DebtStreamness is robust to this class of uncertainty and supports the reliability of our findings, even under limited data availability.

Having established the DebtStreamness framework and introduced the underlying data, we now turn to its empirical application. In the following section, we examine the structure of Uruguay’s inter-firm credit network through the lens of DebtStreamness—first at the firm level, then at the sectoral level—highlighting the emergence of hierarchical credit chains, identifying key intermediaries, and testing the robustness of our findings under alternative network reconstructions.

\section{Results} \label{sec:results}

In ecological systems, trophic levels capture how energy flows through a food web—from primary producers to apex consumers—revealing hierarchical structures and potential ecosystem vulnerabilities. Drawing on this analogy, we now analyze the structure of the Uruguayan inter-firm credit network through the lens of \textit{DebtStreamness}, the economic counterpart to trophic level. By examining both firm- and sector-level positions within the network, we identify patterns in credit flow hierarchies, assess the role of intermediaries, and evaluate how network features such as loops or aggregation affect systemic positioning and potential network fragilities.

\begin{figure}[H]
    \centering
    \caption{\textbf{DebtStreamness of firm network.} DebtStreamness (DS) of firms in the Uruguayan inter-firm credit network. The average value of DebtStreamness is 1.67, with three outliers with $DS>20$.}
    \label{DSfirms}
    \includegraphics[width=10cm, valign=t]{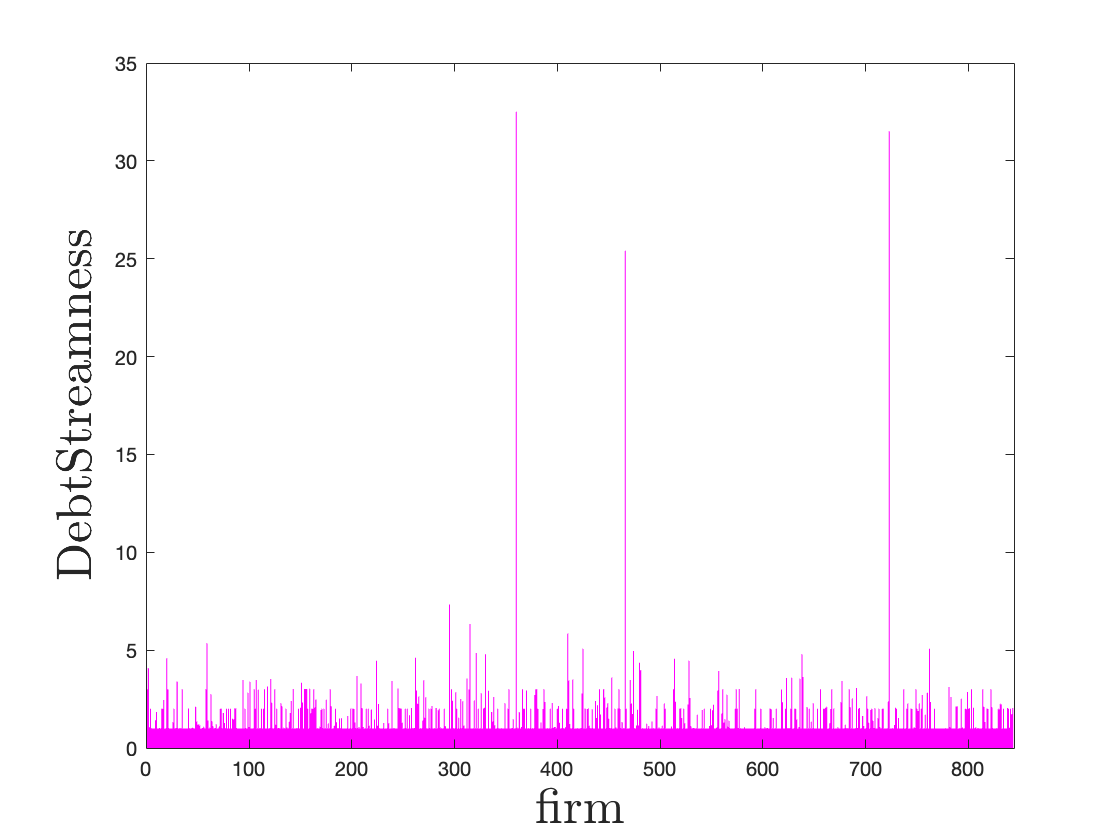}
    
\end{figure}

In Figure \ref{DSfirms}, we show the DebtStreamness of firms within the Uruguayan inter-firm credit network. The average DebtStreamness is $1.67$, indicating that, on average, firms are relatively close to the primary credit source --- the banking sector --- akin to species feeding near the base of an ecological pyramid. The minimum value of DebtStreamness is 1, which corresponds to a firm that borrows from banks only and most firms have Debstreamness values below 6, suggesting rather short credit chains. However, we identify three outliers with DebtStreamness above 20, reflecting their remote, multi-layered positioning within the credit web. These firms are further away from the banking sector along a more complex credit chain.

As mentioned earlier (see Fig. \ref{fig:networkRepresentation} in Sec. \ref{sec:method}), the inter-firm credit network is fragmented into several disconnected components, each representing a self-contained sub-system of inter-firm credit relationships. Within a given component, the DebtStreamness of firms depends solely on the internal structure, since no credit paths link them to firms outside the component, allowing credit transfers from one component to another. This structural isolation resembles ecological sub-networks or isolated food chains, where energy flows are confined within distinct ecosystems.

This separation allows us to analyse each component independently, shedding light on how local network topology influences firms' positions along credit chains. In what follows, we first focus on the largest component, which captures the majority of firms and illustrates the typical hierarchical patterns of DebtStreamness across the broader economy. We, then, turn to a smaller component that contains the three most extreme outliers --- firms positioned farthest from the banking sector.

By contrasting these two cases --- a large, relatively hierarchical component and a smaller, structurally atypical component --- we highlight how fine-grained network features, such as loops, can amplify credit chain lengths and shape systemic risk. This comparative approach highlights the importance of granular, component-level analysis in uncovering hidden vulnerabilities that would otherwise be masked in aggregate statistics.

\subsubsection*{Largest component}
\begin{figure}[htb!]
     \centering
     \caption{\textbf{Left panel:} Histogram of the share of inter-firm credit for firms in the largest component. Approximately $55\%$ of firms primarily rely either on inter-firm credit or direct borrowing from banks, while the rest use a mix of both.
    {\bf Right panel:} Histogram of DebtStreamness values for firms in the largest component. The three peaks suggest a hierarchical structure in the credit chains.}\label{fig:histDeltaDS}

     \begin{subfigure}[b]{0.45\textwidth}
         \centering
         \includegraphics[width=\textwidth]{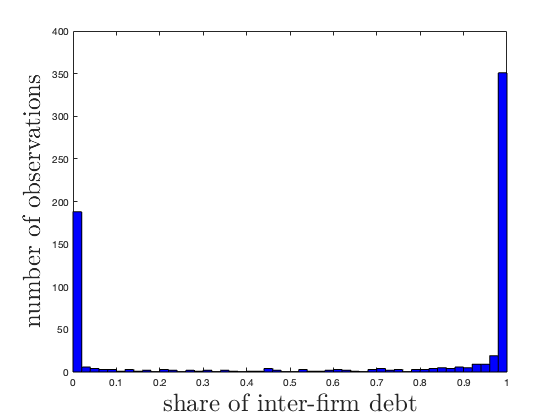}
     \end{subfigure}
     \hfill
     \begin{subfigure}[b]{0.45\textwidth}
         \centering
         \includegraphics[width=\textwidth]{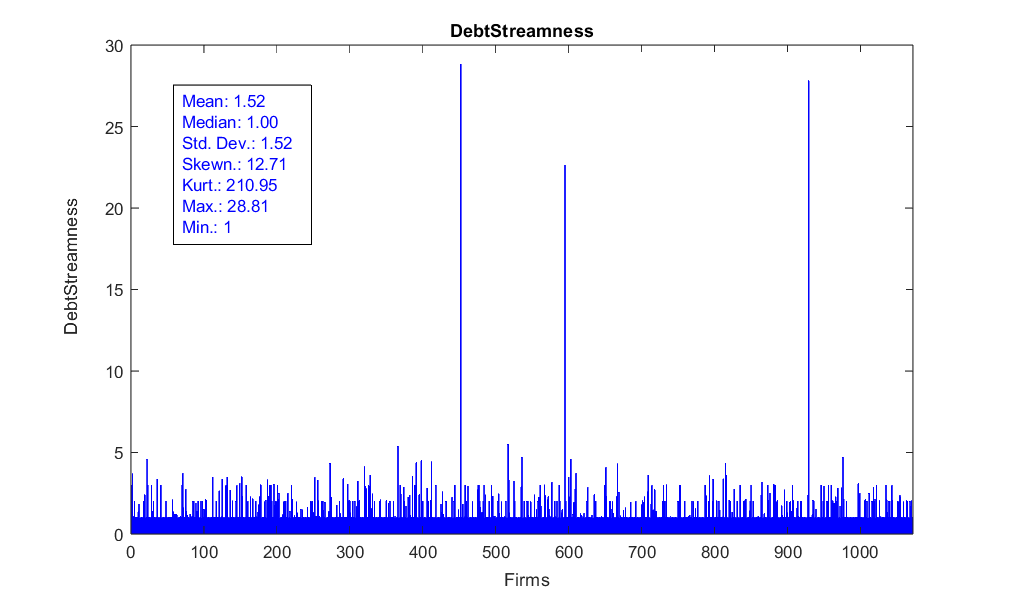}
     \end{subfigure}

\end{figure}

The left panel of Figure \ref{fig:histDeltaDS} shows the histogram of the share of inter-firm debt for nodes in the largest component. The distribution is bimodal: approximately $25\%$ of firms borrow very little from other firms, while around $30\%$ heavily rely on inter-firm debt. The remaining $45\%$ firms are spread between the two extremes. 
These two financing sources, inter-firm credit and bank credit mostly exhibit a substitute relationship, which is in accordance with the results found in \cite{Gopalan2022} and \cite{Petersen2015}. The likelihood of a firm obtaining bank credit may depend on its assets available to secure a bank loan. In cases where a firm lacks sufficient assets or during an economic downturn, it may become difficult for the firm to secure financing from banks \cite{Gopalan2020}. Instead, firms may resort to interfirm credit, suggesting that it acts as a substitute for bank credit when a firm cannot obtain a bank loan. Without alternative funding sources such as banks or investors, firms become more dependent on interfirm credit to continue their operations \cite{Coleman2005}. In \cite{Gopalan2022}, the authors find that in emerging markets and developing economies in Asia, interfirm credit is an important source of financing and serves as a substitute for bank credit.

The distribution of DebtStreamness values is shown in the right panel of Figure \ref{fig:histDeltaDS}. The peaks around 1, 2, and 3 suggest a layered, hierarchical structure in the flow of credit from the banking sector to firms. The highest peak, around 1, confirms that most firms primarily borrow directly from banks. The peak near 2 corresponds to firms that, in turn, borrow significantly from firms with a DebtStreamness of approximately 1. Similarly, the peak around 3 represents firms borrowing from those with a DebtStreamness of about 2.

\begin{figure}[H]
     \centering
     \caption{\textbf{Visualisation of the inter-firm credit network.} Largest component of the inter-firm credit newtork. Nodes represent firms, with colours indicating DebtStreamness ($DS$) values: red nodes have $DS < 1.5$, pink nodes have $1.5 \leq DS \leq 2.5$, and blue nodes have $DS > 2.5$. The banking sector (yellow triangle) serves as the primary originator of credit.}   
     \includegraphics[width=\textwidth]{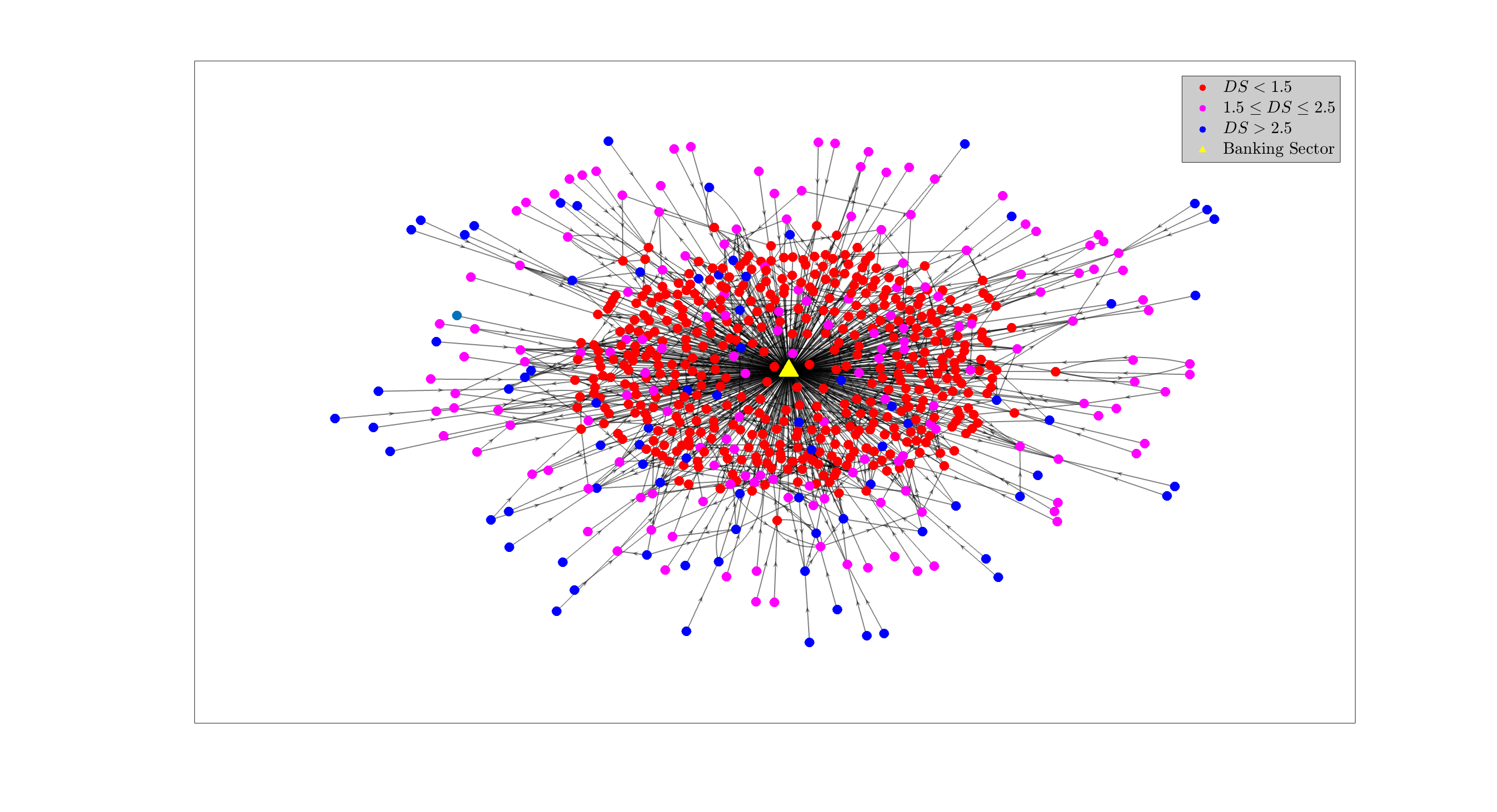}\label{fig:networkHierarchy}
\end{figure}

This stratified structure --- reminiscent of trophic levels in a food web --- is also visually apparent in Figure \ref{fig:networkHierarchy}, where nodes are color-coded by DebtStreamness value. Firms closer to the banking sector (yellow triangle) occupy central, red-colored positions, and have $DS < 1.5$. Firms relying on extended intermediation form the outer layers and populate the peripheral network regions in blue and purple --- much like secondary or tertiary consumers in ecological networks --- showing respectively a DebtStreamness $1.5 \leq DS \leq 2.5$ and $DS > 2.5$.

The structure highlights a tiered flow of credit, with firms closer to the banking sector exhibiting lower $DS$ values, while peripheral firms are associated with higher $DS$ values due to longer credit chains.

\subsubsection*{Component with outliers}
Having analyzed the largest component, where DebtStreamness values exhibit a clear, layered hierarchy, we now turn to a smaller but particularly interesting subnetwork—one that contains the firms with the highest DebtStreamness values in the entire dataset (see Fig. \ref{DSfirms}). Studying this component offers a complementary perspective: it allows us to see how localised structural anomalies, such as extended chains and feedback loops, can dramatically alter a firm's systemic positioning, similarly to how isolated clusters or strong cycles in ecological networks may distort energy flows and destabilize local food webs \cite{allesina2008network}.

\begin{figure}[H]
    \centering
    \caption{\textbf{Top left panel:} Network representation of the component with the three outliers. A node's size represents its total borrowing, and links indicate lender-borrower relationships. Two of the outliers do not borrow directly from the banking sector and form a loop in the network. The third outlier primarily borrows from one of the other two. {\bf Top right panel:} Blue bars (right $y$-axis) show the fraction of a firm's total borrowing that comes directly from the banking sector, while pink bars (left $y$-axis) represent the firm's DebtStreamness value. The three outliers primarily rely on inter-firm credit.
    \textbf{Bottom left panel:} Graph representation of the component with the three outliers after removing the loop between two of them.
    {\bf Bottom right panel:} Comparison of DebtStreamness values in the original component (pink bars) versus the modified component with the loop removed (green bar). Removing the loop significantly reduces the length of credit chains in the system.
    }\label{fig:Component1bankoutlier}
    \begin{subfigure}[b]{0.35\textwidth}
    \centering
      \includegraphics[width=\textwidth]{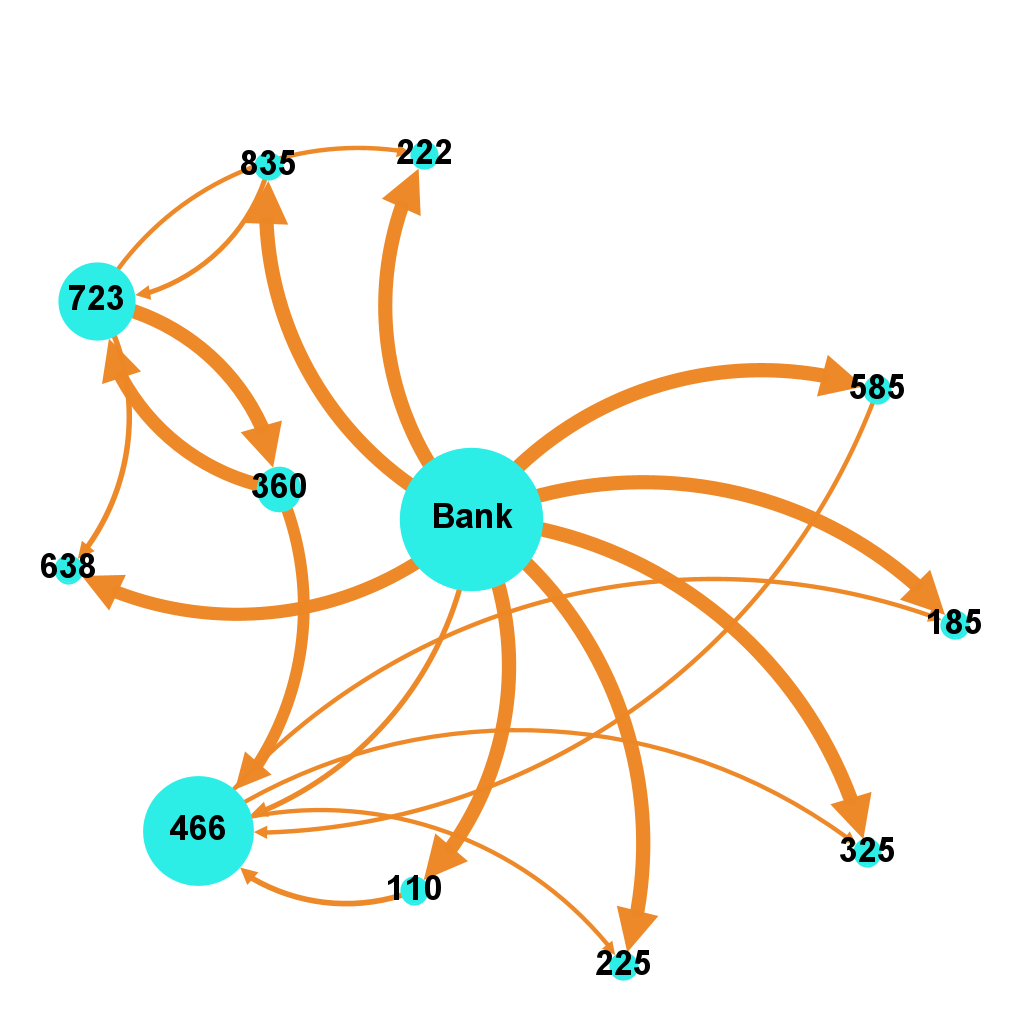}    \end{subfigure}%
    \begin{subfigure}[b]{0.35\textwidth}
    \centering
    \includegraphics[width=\textwidth]{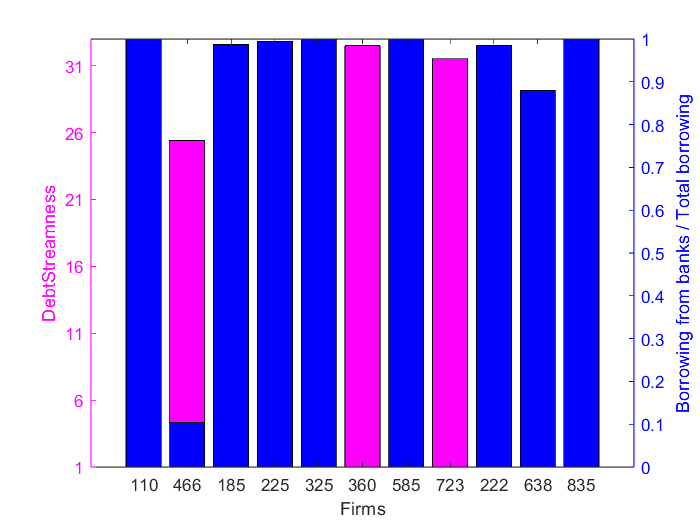} 
    \end{subfigure}
    \begin{subfigure}[b]{0.35\textwidth}
    \centering
    \includegraphics[width=\textwidth]{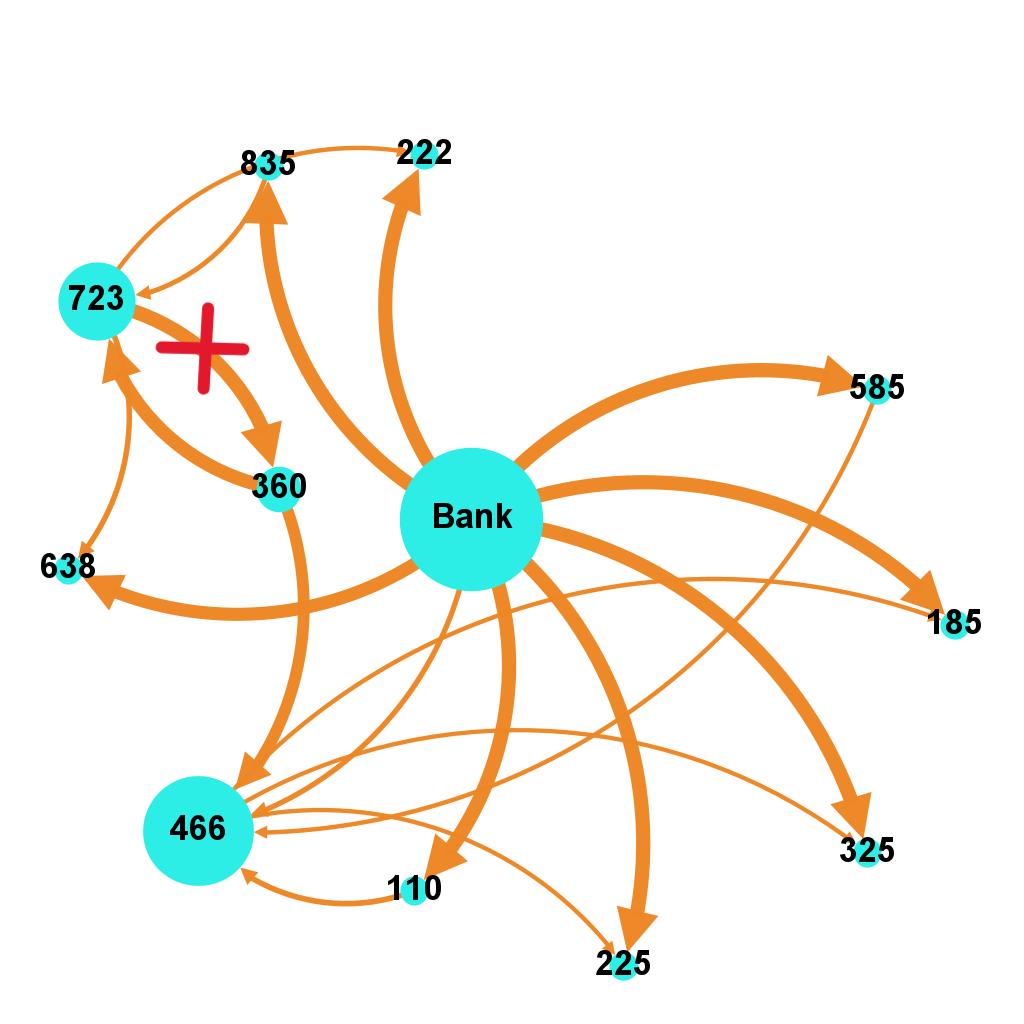}
    \end{subfigure}%
    \begin{subfigure}[b]{0.35\textwidth}
    \centering
     \includegraphics[width=\textwidth]{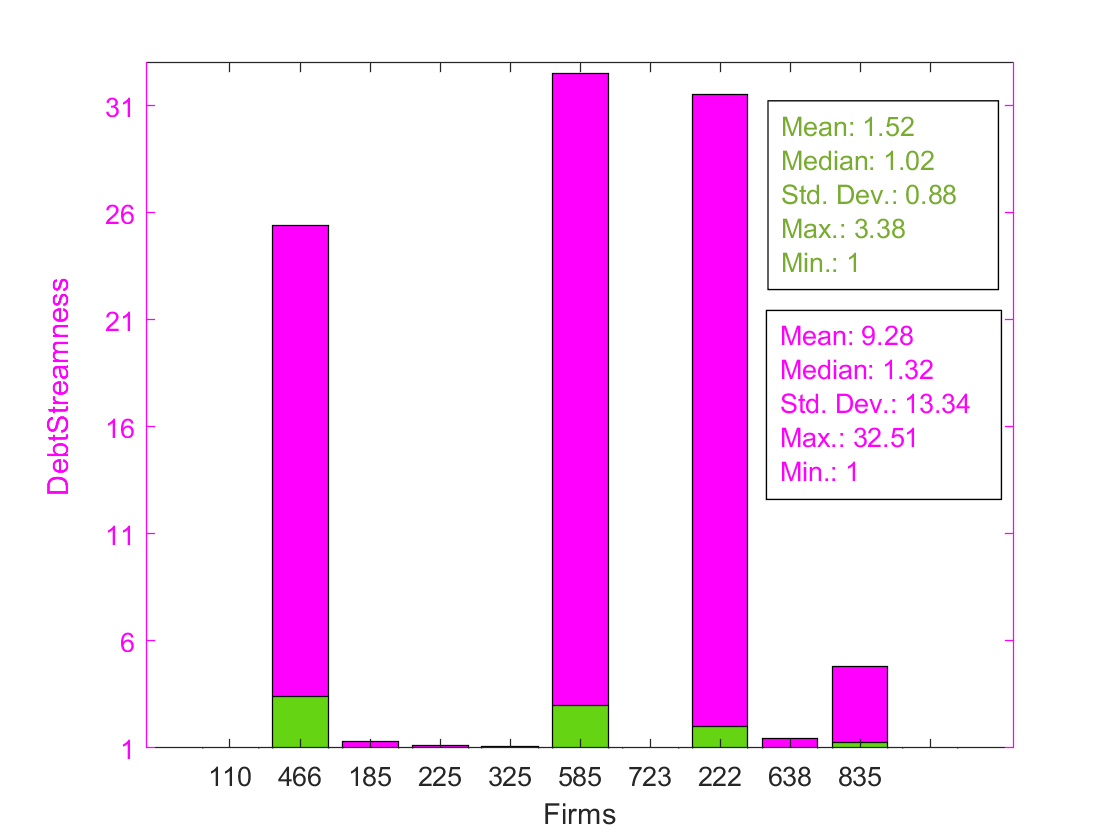}
    \end{subfigure}\label{Componentoutlier}
\end{figure}

The top left panel of Figure \ref{fig:Component1bankoutlier} shows the component with the three outliers and illustrates their connections in the network. The direction of arrows in the figure goes from the lender to the borrower, and a representative node of the banking system is included at the center. We see that two of the outliers (nodes $360$ and $723$) are not directly connected to the banking sector, while the third (node $466$) mostly borrows from node $360$. Additionally, the two outliers with no direct borrowing from the banking sector are involved in a loop of length two, reinforcing credit circulation between them and increasing their DebtStreamness. Two main factors contribute to their increased DebtStreamness: i) paths from the banking sector to these three firms are longer due to intermediaries, and ii) the sum in Equation \ref{eq:DebtStreamnessDistance} will contain infinite terms, corresponding to paths that repeatedly go through the loop.

In the remaining panels of Figure \ref{fig:Component1bankoutlier}, we explore these two aspects separately. In the top right panel of Figure \ref{fig:Component1bankoutlier},  we plot for each firm in the component its DebtStreamness (pink bars) and the fraction of its debt borrowed directly from the banking sector (blue bars).  We observe a strong negative correlation between the two quantities, which is around $-0.99$. This seems to suggest that, for the system at hand, the network structure is not that important; rather, in agreement with the results of \shortciteA{Bartolucci2020, bartolucci2025upstreamness, Bartolucci_2021} obtained for related centrality measures, knowing the fraction of money that a firm directly borrows from the banking sector is enough to estimate, on average, its DebtStreamness with good approximation. However, we will now see that certain details of the network can make a significant difference in terms of the DebtStreamness of specific nodes.
To this end, we compute the change in DebtStreamness of nodes in the component that would occur if we killed one loop by removing the link from node $723$ to node $360$. This is shown in the bottom panels of Figure \ref{fig:Component1bankoutlier}. By removing the loop, we observe a significant reduction in DebtStreamness, with an average DebtStreamness of the component that decreases from $9.28$ to $1.52$.  The presence of feedback loops can dramatically shift a firm’s position within the credit hierarchy: this result highlights the disproportionate impact of small topological motifs on systemic positioning, echoing findings from ecological stability studies where feedback loops are known to critically influence the robustness and resilience of food webs \cite{neutel2002stability}.

\subsection{Sector analysis}

While firm-level analysis reveals a fine-grained credit hierarchy, aggregating firms by sector allows us to explore whether such ``trophic-like structures" persists at higher organizational levels.

To perform sectoral analysis, starting from the matrix $L_{ij}$ of inter-firm credit, we construct the aggregate matrix $L^{\rm (s)}$, where $L^{\rm (s)}_{kt}=\sum_{i\in k}\sum_{j\in t} L_{ij}$ is the aggregate amount borrowed by firms of sector $k$ from firms of sector $t$, and we define the total debt of sector $k$ as $D^{\rm (s)}_k=\sum_{i\in k} D_i$ . After this aggregation, we can proceed as before to define the DebtStreamness $DS^{\rm (s)}_k$ of sector $k$ as
\begin{equation}
    DS^{\rm (s)}_k = 1 +\sum_t A^{\rm (s)}_{kt} DS^{\rm (s)}_t \ ,
\end{equation}
where $A^{\rm (s)}_{kt}=L^{\rm (s)}_{kt}/D^{\rm (s)}_k$ \ .

\begin{figure}[H]
    \centering
    \caption{\textbf{Sector aggregation.} \textbf{Left Panel}: $DS$ and bank-debt ratio of sectors. \textbf{Right Panel} Network component visualisation aggregated by sector with the bank node as originator.} \label{fig:Sectors}
    \begin{subfigure}[b]{0.5\textwidth}
    \centering
    \includegraphics[width=6.5cm, valign=m]{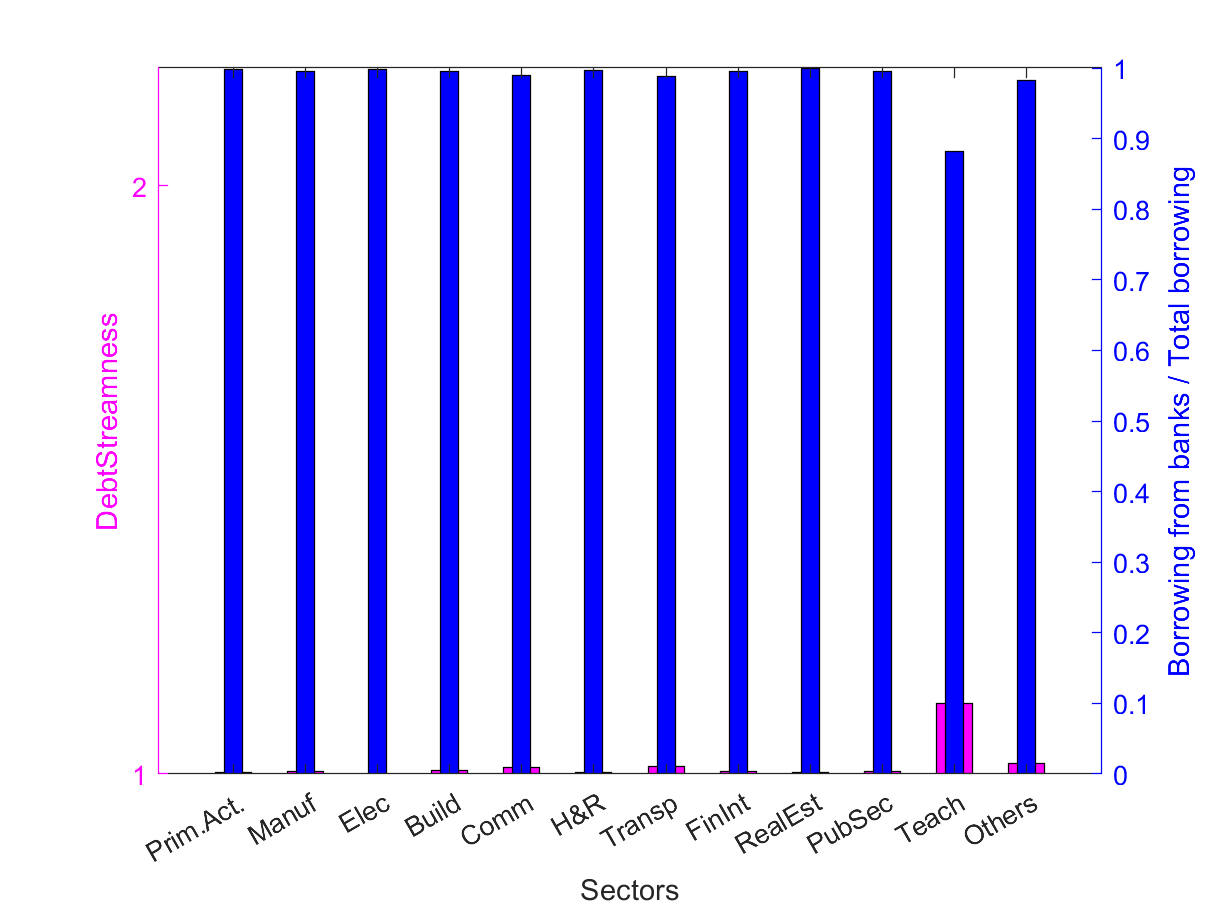}
    \end{subfigure}%
    \begin{subfigure}[b]{0.5\textwidth}
    \centering
    \includegraphics[width=7cm, valign=m]{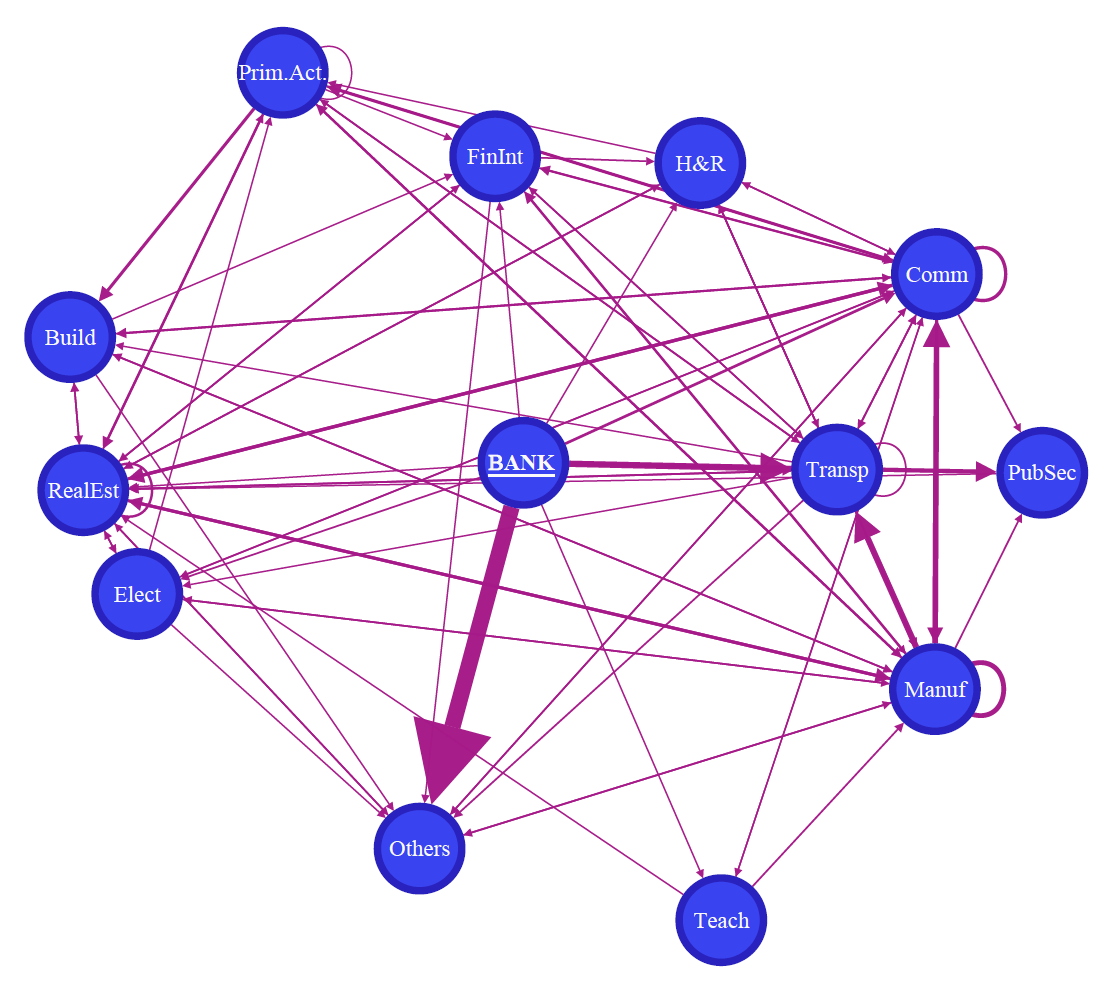}
    \end{subfigure}
\end{figure}

The list of sectors in our dataset, number of firms, and the amounts borrowed by each sector from banks and firms is reported in Table \ref{table:Sector}\footnote{Firms in the primary activities, real estate, and financial intermediation sectors appear only as debtors or creditors of the surveyed firms.}, and the results of this aggregation are shown in Figure \ref{fig:Sectors}.

In the left panel of Figure \ref{fig:Sectors}, we show the DebtStreamness alongside the share of borrowing from banks for each sector. In the right panel, we have the corresponding sectoral-level network representation with the bank node as credit originator. The figure shows that most sectoral borrowing comes from banks rather than inter-firm relationships, resulting in DebtStreamness values close to 1 across all sectors. This pattern is also evident in the right panel, where all sectors are directly connected to the node representing the banking system. The trivial credit chain structure at the sectoral level suggests that key interfirm credit relationships are obscured in aggregated data, highlighting the importance of fine-grained data to accurately capture the underlying economic dynamics.

A similar effect is detected in ecology, where fine-grained food web studies are needed to detect keystone species or hidden energy pathways \cite{bascompte2005simple, abarca2002effects}.

While our DebtStreamness analysis captures how far sectors are positioned from the primary source of credit (the banking sector), it is also insightful to compare this financial hierarchy with more traditional classifications of economic structure. In particular, sectoral roles within production networks—whether primarily input suppliers (upstream), final good producers (downstream), or key intermediaries—may influence their borrowing patterns and exposure along credit chains.

To make this comparison concrete,  we adopt the classification of sectors based on the Uruguayan economic structure analysis from the Central Bank of Uruguay  \cite{BCU2023}, and we grouped the sectors into three categories: Upstream, Key sectors, and Downstream. We labeled as ``Others" the nodes for which sector classification is unknown. In this classification, sectors are categorized using the Rasmussen methodology \cite{rasmussen1956studies} or the Backward ($BL$) and Forward Linkages ($FL$) approach. The $BL$ measures how much a sector depends on its suppliers, while the $FL$ measures how much a sector supplies inputs to other sectors. If $BL > 1$, it indicates that sector $j$ has a stronger backward linkage than the economy's average, meaning it is heavily dependent on other sectors for inputs. Similarly, if $FL > 1$, it indicates that sector $i$ has a stronger forward linkage than the economy's average, implying that many downstream firms depend on it. Based on these linkages, if $BL > 1$ and $FL < 1$, the sector is classified as upstream, as it provides inputs to other sectors but do not have strong downstream dependencies. If $BL < 1$ and $FL > 1$, the sector is classified as downstream, as it has a strong dependence on upstream sectors but contributes less to their input demands. Finally, if $BL > 1$ and $FL > 1$ , the sector is considered a key sector, meaning it plays a crucial role in both directions, serving as a both major supplier and consumer in the economy.

This structural classification provides an independent perspective on economic positioning, allowing us to assess whether a sector’s DebtStreamness --- its financial distance from primary credit sources --- aligns with its traditional production role, or reveals hidden vulnerabilities not visible through production linkages alone.

\begin{figure}[H]
     \centering
     \caption{\textbf{Sector classification by DS.} This plot represents the sector classification of firms and their DebtStreamness ($DS$), categorized into UpStream, Key sector, DownStream, and Others with no classification available across three $DS$ ranges: $DS < 1.5$, $1.5 \leq DS \leq 2.5$, and  $DS > 2.5$.}   
     \includegraphics[width=0.8\textwidth]{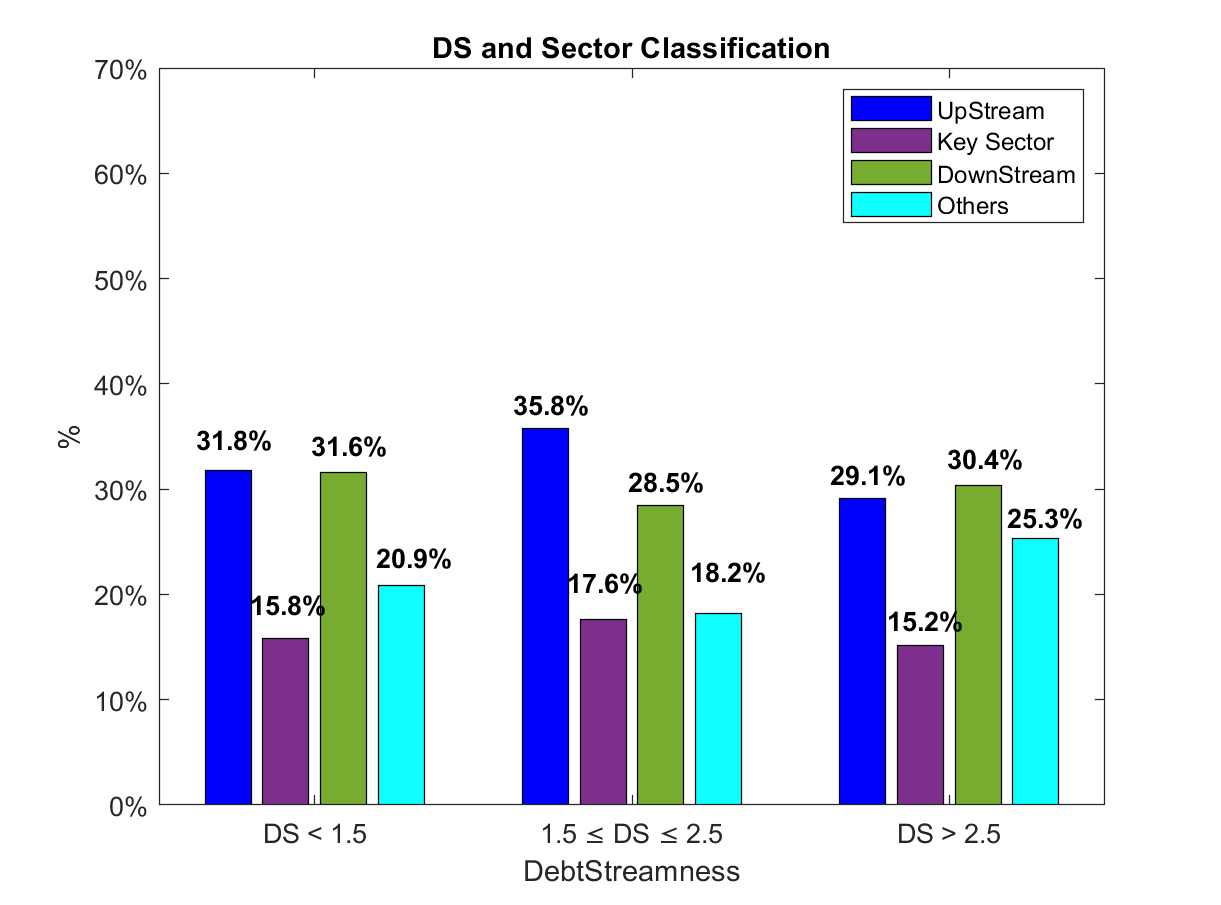}\label{fig:DSclassbar}
\end{figure}

Figure \ref{fig:DSclassbar} shows how sector classifications—based on production linkages—map onto firms’ positions within the credit hierarchy, as captured by DebtStreamness. Across the three DS ranges ($DS < 1.5$, $1.5 \leq DS \leq 2.5$, and $DS > 2.5$), upstream and downstream sectors appear in roughly equal proportions. However, the middle DS range ($1.5 \leq DS \leq 2.5$) contains a noticeably higher share of upstream sectors, suggesting that these firms often combine both bank loans and inter-firm credit. When $DS > 2.5$, the downstream sector is slightly more likely to obtain financing from other firms and is further from the original credit source.

This result suggests that upstream sectors may enjoy more diversified or proximate access to credit, while downstream sectors are relatively more dependent on cascading inter-firm lending—farther from the original source of financial energy. This finding reinforces the value of DebtStreamness as a complementary lens to production-based classifications, revealing differences in financial positioning not captured by input-output linkages alone.

\begin{table}[h]
    \centering
    \renewcommand{\arraystretch}{1.2}
    \begin{tabular}{l l l l l}
\hline
        \textbf{Sector} & \textbf{Nodes} & \textbf{Ratio of firm's} \\ 
        & & \textbf{borrowing from banks} \\  
        & & \textbf{to its total borrowing (\%)} \\ 
\hline
Primary Activities & 64 & 18.9\%  \\
        Manufacturing & 267 & 31.7\%  \\
        Electricity & 9 & 3.4\%  \\
        Building & 32 & 8.2\%  \\
        Commerce & 297 & 15.6\%  \\
        Hotels \& Restaurants & 17 & 1.3\%  \\
        Transportation & 104 & 8.3\%  \\
        Financial Intermediation & 7 & 0.2\%  \\
        Real Estate & 2 & 1.2\%  \\
        Public Sector & 29 & 5.4\%  \\
        Teaching & 15 & 0.0\%  \\
        Others & 229 & 5.7\%  \\
\hline        \textbf{Total} & 1072 & 100.0\% \\
\hline
    \end{tabular}
    \caption{Sectoral aggregate firm credit information.}
    \label{table:Sector}
\end{table}

\subsection{Robustness analysis: partially reconstructed networks} \label{sec:reconstruction}

\noindent

So far, our analysis has been based on the network constructed by linking each surveyed firm to its top three creditors and borrowers. However, as described in Sec. \ref{sec:method}, we have also access to survey data regarding information on each firm's \textit{total volume} of inter-firm credit, allowing us to estimate the extent of missing links in the observed network.

As shown in Figure~\ref{fig:3firms}, while top three creditors account on average for about 50\% of a firm's total inter-firm credit, the distribution is highly heterogeneous across firms. This naturally raises the question of how the residual, unreported credit relationships might affect the DebtStreamness values we have computed.

To address this, and following the reconstruction procedures introduced in Section \ref{sec:method}, we conduct robustness checks by constructing two alternative partially reconstructed networks: 
\begin{itemize}
    \item A fully connected reconstruction, where each firm's residual credit is spread uniformly across all non-top-three partners.
\item A sparse reconstruction, where residual credit is assigned to a minimal number of partners, maintaining consistency with observed credit intensities.
\end{itemize}.

These reconstructed networks allow us to test whether the structure of missing credit relationships could effectively distort the systemic positioning of firms as measured by DebtStreamness.

\begin{figure}[H]
    \centering
    \caption{{Comparison of DebtStreamness computed from the network with only top three creditors and the partially reconstructed ones. {\bf Left panel}: comparison with partially reconstructed fully connected network. {\bf Right panel}: comparison with partially reconstructed sparse network.}}
    \label{correlationmet1}
    \includegraphics[width=7cm]{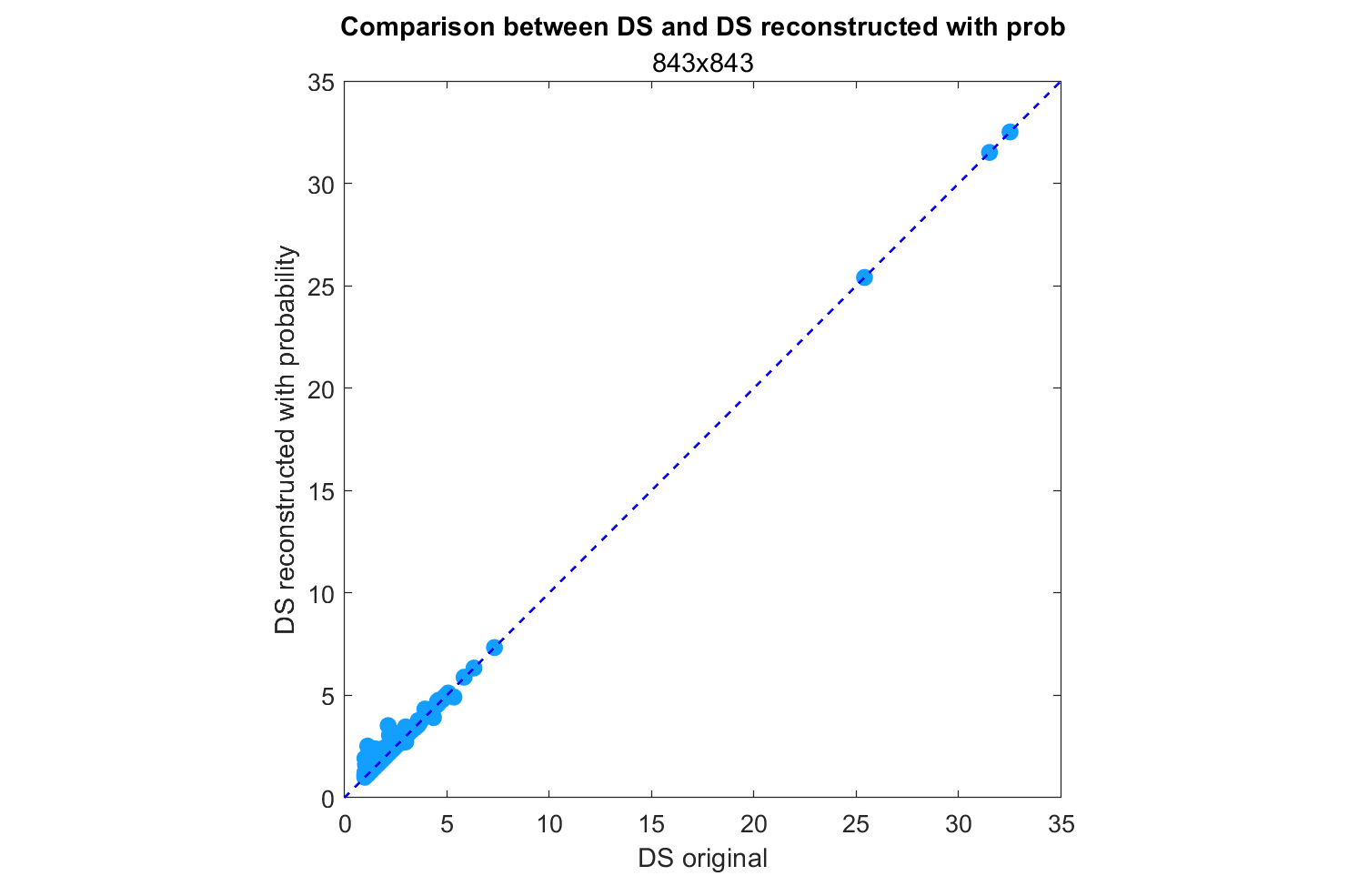}
        \includegraphics[width=6.2cm]{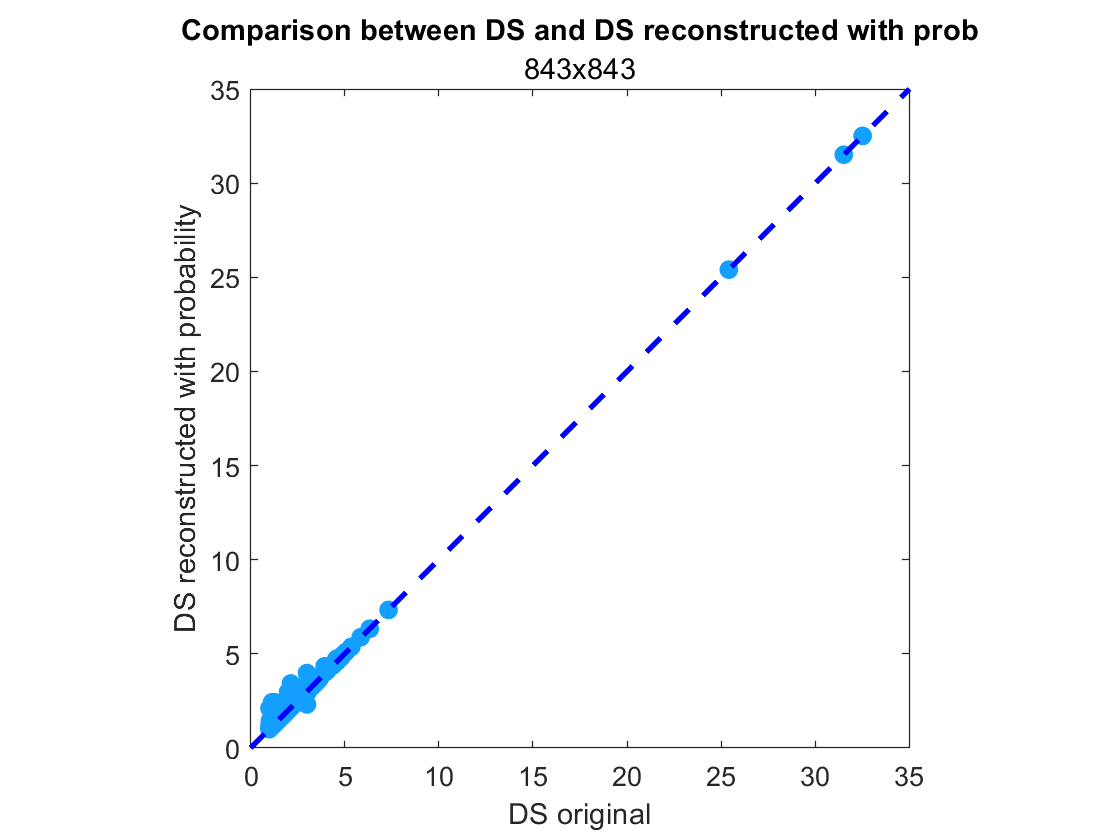}
\end{figure}

The results, reported in Figure~\ref{correlationmet1}, show the original DebtStreamness values (computed from the observed top-three debtors/creditors network, $L^{\text{top}}$) against those obtained from the two reconstructed networks. In both cases, we observe extremely high correlations: Spearman correlations are $0.99$ for both full and sparse networks, and Kendall correlations are equal to $0.95$ and $0.96$ for full and sparse network respectively.

This high degree of robustness is consistent with two key observations: (i) top creditors already account for a substantial share of inter-firm credit, and (ii) as shown earlier, firms are generally located close to the banking sector in the credit hierarchy, limiting the potential for long, complex chains to form through the missing links.

Overall, the strength of the agreement across reconstructed networks confirms that our findings on DebtStreamness --- and the associated insights into credit chain structure and systemic positioning --- are not sensitive to the partial nature of the available data. This reinforces the practical value of the DebtStreamness metric for economic network analysis even in settings with incomplete information, a common feature in real-world data collection.

\section{Conclusions} \label{sec:conclusions}
Drawing inspiration from ecology, this work develops an interdisciplinary approach to studying financial networks by adapting concepts from food web analysis ---specifically, the idea of trophic levels --- to the flow of credit between firms. In ecological systems, trophic levels capture how energy cascades from primary producers through successive layers of consumers; similarly, we conceptualize financial systems as structures in which credit originates from banks and flows through firms via inter-firm lending relationships.

Building on this analogy, we introduced the metric of DebtStreamness to quantify the hierarchical structure of credit chains within an economy. DebtStreamness measures a firm's average distance from the banking sector --- the origin of financial ``energy'' --- across all possible credit paths in the inter-firm network. Our measure provides a new tool for mapping financial dependencies and systemic positioning in a manner that complements traditional production network analyses.

As a use case, we applied our metric to the analysis of interfirm credit in Uruguay, using data from the 2018 Economic Expectations Survey conducted by the Central Bank of Uruguay. Our analysis uncovers the following three key findings. First, credit chains are generally short: the average DebtStreamness is close to 1.7, indicating that most firms rely predominantly on direct bank financing. However, we also observe a tiered structure, where some firms borrow primarily from banks and, in turn, provide credit to others. Second, local network motifs --- notably feedback loops --- can inflate DebtStreamness dramatically for specific firms, creating hidden pockets of increased systemic exposure despite similar levels of direct bank borrowing. Third, in spite of substantial missing data on smaller credit relationships, DebtStreamness remains highly robust under alternative network reconstructions, demonstrating its practical applicability when only partial information is available.

Moreover, DebtStreamness offers insights that go beyond conventional production‐based classifications of firms as upstream or downstream.  These findings could support financial regulators and policymakers in identifying firms that act as hidden intermediaries in the credit network --- entities that may not appear systemically important based on size alone, but whose network position makes them critical in the transmission of financial stress.

Although recent events such as the COVID-19 pandemic and the global energy crisis have drawn renewed attention to the fragility of supply chains and economic interdependencies, our findings show that financial flows—and specifically credit relationships—constitute an additional layer of vulnerability and resilience that deserves closer attention.

Our analysis represents an initial step toward a more comprehensive understanding of inter-firm credit relationships. Future research should investigate different systems and make use of more complete datasets. Additionally, studying how shocks propagate through these networks is vital for assessing the systemic stability of economic systems. A promising direction for future work is the joint analysis of credit and supply chain networks as a multi-layered structure, enabling a fuller understanding of the interdependencies that shape firm behavior and systemic risk.

Ultimately, bridging ecological thinking and economic network analysis may offer valuable new tools for anticipating, managing, and mitigating systemic risks in an increasingly interconnected world.

\section*{Acknowledgments} P.V. acknowledges support from UKRI Future Leaders Fellowship Scheme (No. MR/X023028/1). F.C. acknowledges support of the Economic and Social Research Council
(ESRC) in funding the Systemic Risk Centre at the LSE (ES/Y010612/1).

\bibliography{biblio}

\appendix
\section{Debstreamness derivation} \label{app:DebstDerivation}
DebtStreamness ($DS_i$) --- as defined in \eqref{eq:DebtStreamnessDistance} --- is equivalent to the notion of \emph{downstreamness} in the economic analysis of Input/Output system \cite{bartolucci2025upstreamness, Silvia2023, Antras2012}, this time computed within the inter-firm network using the banking sector as reference point. $DS_i$ represents firm $i$'s average downstream position from the banking sector, which acts as the initial lending node in the interfirm credit network within the economy (see Fig. \ref{fig:DSscheme}). Specifically, Eq. \eqref{eq:DebtStreamnessDistance} defines $DS_i$ for firm $i$ as the value of gross debt of firm $i$, positioned downstream (in the final stage of debt within the production process) with respect to banks, which are positioned upstream and serve as the initial nodes. 

\noindent
By re-summing \eqref{eq:DebtStreamnessDistanceSum} explicitly, we can rewrite it as

\begin{equation}\label{M8}
DS_{i}=\frac{\left[(\mathds{1}-\mathcal{L})^{-2} \overrightarrow{B}\right]_i}{D_i}\ ,
\end{equation}
where $\left[\cdot\right]_i$ denotes the $i$th entry of the vector, $\mathcal{L}=(\ell_{ij})$, and $\overrightarrow{B}$ is the vector of the amounts borrowed by firms $i$ from the banking sector.

\noindent
Inserting now Eq. \eqref{eq:identityTotalDebt}, which can be rewritten as
\begin{equation}\label{eq:identityTotalDebtAppendix}
D_i = B_i + \sum_{j=1}^N L_{ij}= B_i + \sum_{j=1}^N \ell_{ij}D_j\Rightarrow \overrightarrow{D}=(\mathds{1}-\mathcal{L})^{-1}\overrightarrow{B}\ ,
\end{equation}
into Eq. \eqref{M8}, we have the following

\begin{equation}\label{M9}
DS_{i}=\frac{\left[(\mathds{1}-\mathcal{L})^{-1} \overrightarrow{D}\right]_i}{D_i}
\end{equation}

\noindent
where $\overrightarrow{D}$ is the $N \times 1$ vector representing the total amount of debt of firms. 

Introducing the matrix
\begin{equation}
    \Delta = \left(
    \begin{array}{cccc}D_{1} & 0 &\cdots & 0 \\ 0 & D_{2} & \cdots & 0 \\ \vdots & \vdots & \ddots & \vdots \\ 0 & 0 & \cdots & D_{N}\end{array}\right)\ ,
\end{equation}
and recalling that $\ell_{ij}=L_{ij}/D_j$ and $A_{ij}=L_{ij}/D_i$ which implies $\mathcal{L}=L \Delta^{-1}$, $A=\Delta^{-1}L$ and therefore $\mathcal{L}=\Delta A\Delta^{-1}$, we can write from \eqref{M9}

\begin{equation}
 \overrightarrow{D S}=\Delta^{-1}(\mathds{1}-\Delta A\Delta^{-1})^{-1}\Delta \overrightarrow{1}\ .  
\end{equation}
This expression can be further simplified by writing $\mathds{1}=\Delta\Delta^{-1}$ and after simple manipulations as

\begin{equation}\label{M10}
\overrightarrow{D S}=(\mathds{1}-A)^{-1} \overrightarrow{1}\ .
\end{equation}

\noindent
Here, $(\mathds{1}-A)^{-1}$ is the analog of the Leontief-inverse matrix in Input-Output economics. We recall that the matrix $A$ has elements $A_{ij}=L_{i j} / D_{i}$, each representing the share of the borrowing by firm $j$ from firm $i$ divided by the total debt of firm $i$ (input-upstream matrix), while $\overrightarrow{1}$ is the vector $N \times 1$ of ones.

\end{document}